\documentclass[a4paper]{jpconf}
\usepackage{enumerate}
\usepackage{graphicx,epsfig}

\def\beq{\begin{equation}}
\def\eeq{\end{equation}}
\def\bea{\begin{eqnarray}}
\def\eea{\end{eqnarray}}
\def\benu{\begin{enumerate}}
\def\eenu{\end{enumerate}}
\def\nn{\nonumber}
\def\l{\left}
\def\r{\right}

\def\p{\prime}

\begin{document}

\title{Effect of dark energy sound speed and equation of state on CDM power spectrum}

\author{Rizwan Ul Haq Ansari and Sanil Unnikrishnan,}

\address{IUCAA, Post Bag 4, Ganeshkhind, Pune 411 007, India.}

\ead{rizwan@iucaa.ernet.in}

\date{}

\begin{abstract}
We study the influence of equation of state  $w$ and effective sound speed  $c_e$ of the dark energy perturbations on the cold dark matter(CDM)  power spectrum.
 We consider  different cases  of the equation of state  and the effective sound speed, the cold dark matter power spectrum is found to be generically suppressed in
 these cases as compared to the $\Lambda$CDM model. The suppression at different length scales depends on the value of $w$ and $c_e$, and the effect of different $w$ is profoundly seen at all length scales.  The influence of  sound speed is significantly seen only at the intermediate length scales and is negligible at scales very much larger and smaller than the Hubble scale.
\end{abstract}

Recent Type Ia supernova observations have confirmed that the universe is expanding
at an accelerated rate at the present epoch \cite{Reiss}. Understanding the nature of the component which drives this accelerated expansion is one of the major challenges of present day cosmology .  One of the possible candidate could be a dynamical scalar field, with a negative pressure, known as dark energy (DE).
 Several models of scalar field dark energy
 like the quintessence, k-essence, chaplygin
gas, etc  have been studied extensively \cite{Sahni-2004}-\cite{Li-2011}.  In general these models can lead to degenerate evolution of the scale
factor \cite{Paddy-2002}. The scalar field models of dark energy are described by the two parameters
equation of state (EOS) $w$ and speed of sound (ESS)  $c_e$. Since dark energy is a scalar field it not only contributes in the background, but also through the perturbations, so we need to take into account perturbations to the dark energy \cite{Putter-2010,DeDeo-2003}.  The ESS which
relates the density perturbations to pressure perturbation is
useful in studying the effects of perturbations in DE. So, considering the DE perturbations can provide vital information in understanding these models, hence
distinguish and possibly ruling out some of the models. Our aim here is to study how perturbation in such models
of dark energy, will effect the cold dark matter
 power spectrum. We will focuss on evolving equation of state and speed of sound, whose evolution is not fixed by a particular model of DE, and see  its
effects on observable quantities such as CDM power
spectrum.
\begin{figure}[t]
{\includegraphics[scale=0.35]{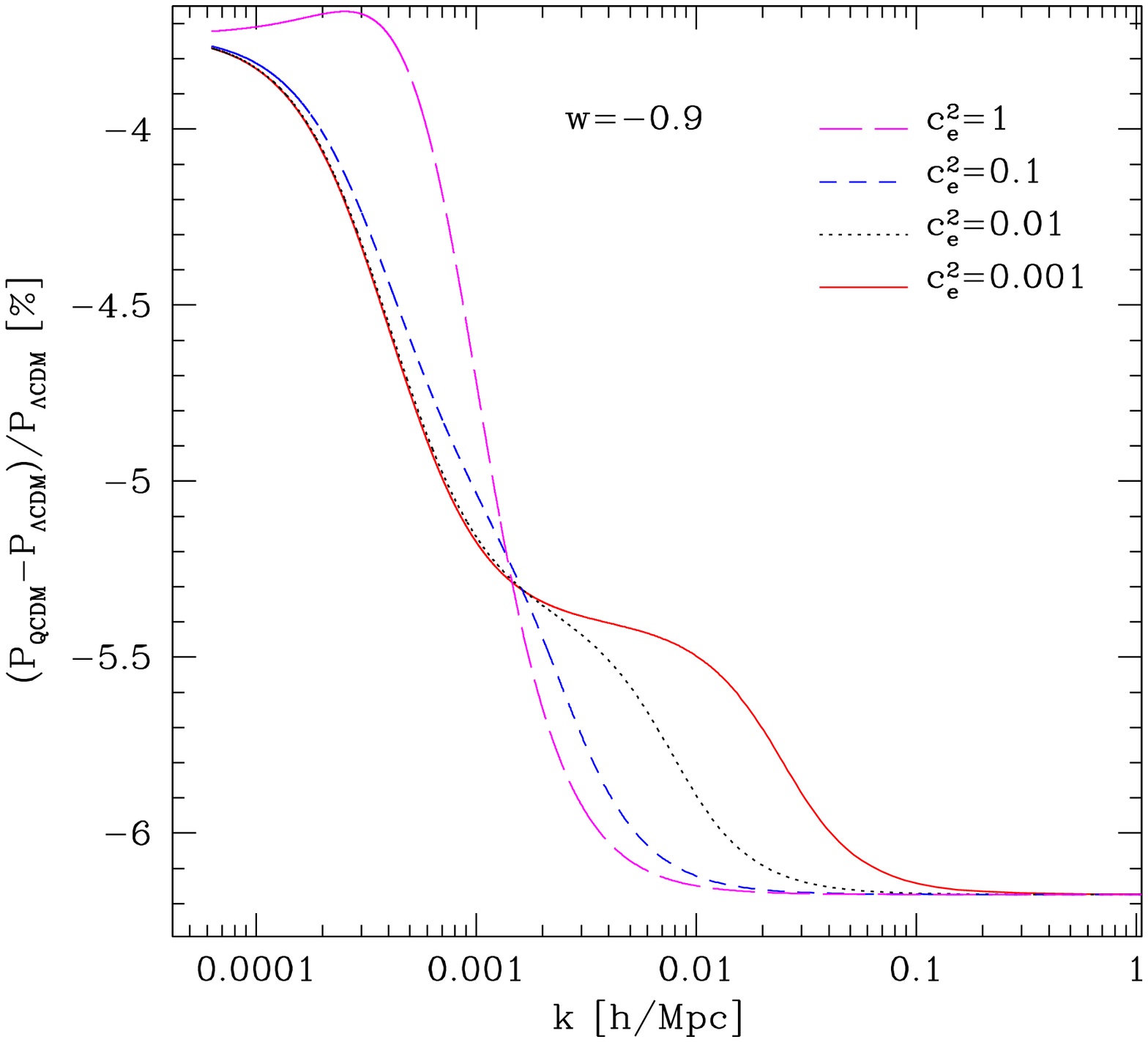}}
{\includegraphics[scale=0.35]{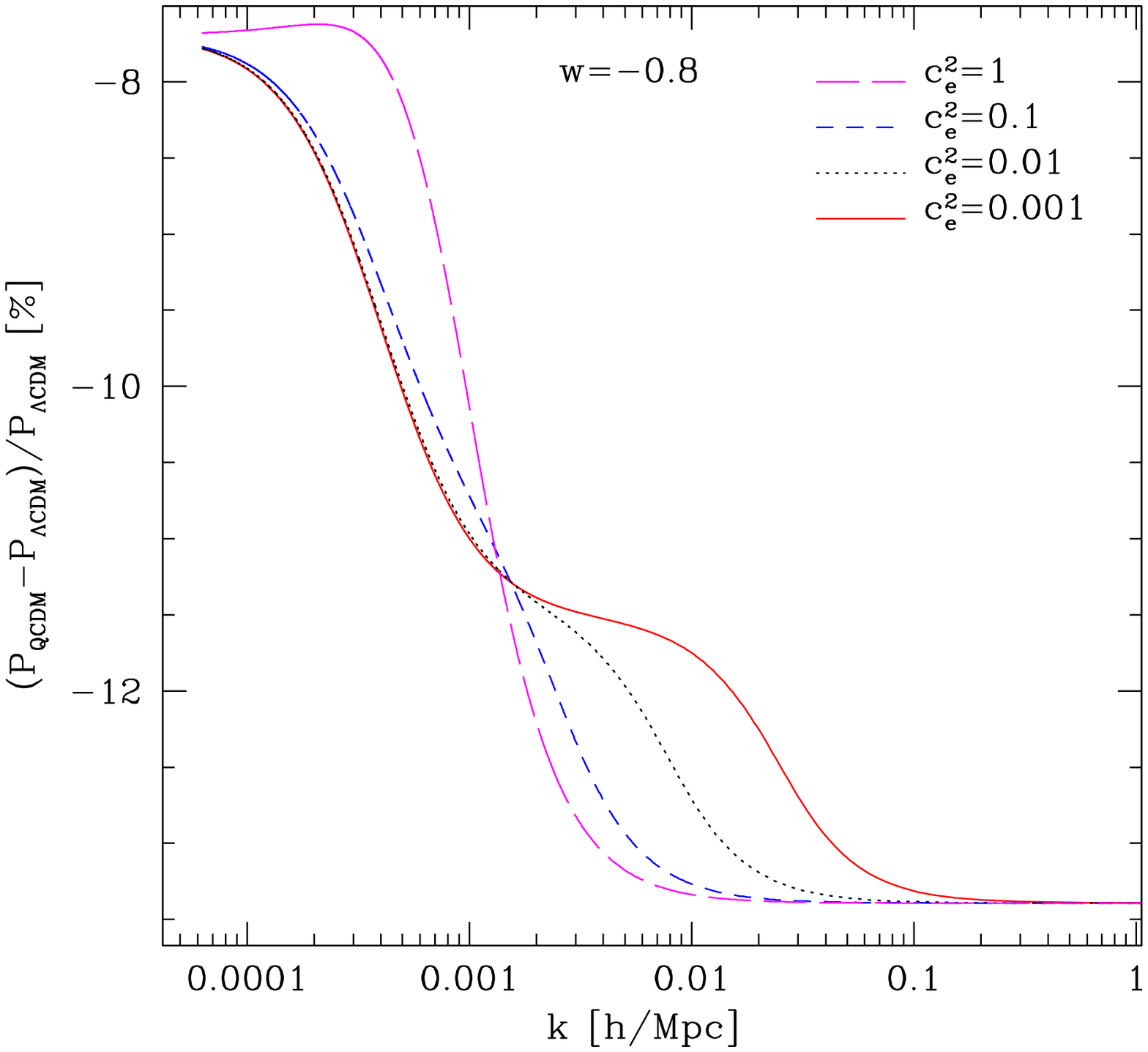}} \caption{The
percentage suppression in matter power spectrum in dark energy
models in comparison to $\Lambda$CDM model is plotted for a
constant $w$ and for different values of
$c_e^2$.  It can be seen that suppression is more
for dark energy  models with  $w=-0.8$ than with $w=-0.9$.
The power spectrums shown in this figure corresponds to its value
at the present epoch.}\label{F1}
\end{figure}

We consider a scalar field model of dark energy, whose action is
given by: $$S[\phi] = \int\  \sqrt{-g}\; {\cal L}(X,\phi) \,
d^{4}x,$$ where ${\cal L}(X,\phi)$ is the Lagrangian density of the
scalar field. Also we assume the universe to be spatially flat consisting of cold dark matter and dark energy. The equation of state
parameter for the dark energy with above mentioned action is given
by $ w=\frac{\mathcal{L}(X,\phi)}{2\l(\frac{\partial
\mathcal{L}}{\partial X}\r) X - \mathcal{L}(X, \phi)}$, the
corresponding effective speed of sound of DE is $c_{e}^{2} \equiv
\frac{\l(\frac{\partial \mathcal{L}}{\partial X}\r)}{\l(\frac{\partial \mathcal{L}}{\partial X}\r)
+ 2
 X\l(\frac{\partial^{2}\mathcal{L}}{\partial X^{2}}\r)}$. From these definitions of EOS and ESS, it is obvious that there evolution is governed by the Lagrangian density of the scalar field dark energy. However, it is also possible to obtain the form of Lagrangian density from a given evolution of $w$  parameter and $c_e^2$  (see \cite{Rizwan}). Here, we will use this fact and choose a parameterized form of $w$ and $c_e^2$ to study the influence of perturbations in DE on CDM power spectrum.
We start by presenting the necessary equations for the study of evolution of perturbations, the perturbed
Friedmann-Robertson-Walker
 (FRW) metric in the
longitudinal gauge is,
\begin{equation}
    ds^{2} = \left( 1 + 2\Phi\right)dt^{2} - a^{2}(t)\left(1 - 2 \Phi\right)\left[dx^2 + dy^2 +
    dz^2\right], \label{eqn::longitudinal gauge}
\end{equation}
where  $\Phi$ is the gravitational potential describing the scalar metric perturbations.
In a system consisting  of CDM and dark
energy, using the linearized Einstein's equations and the covariant
conservation equations for dark energy, we can obtain following set of equations for the evolution of perturbations:
\begin{eqnarray}
\Phi^{\p\p} &=& - \frac{\Phi^{\p }}{a}\left(
\frac{\ddot{a}}{aH^2}+4 \right)- \frac{\Phi}{a^2}\left(
2\frac{\ddot{a}}{aH^2}+ 1 \right) + \frac{4 \pi G \bar{\rho}_{d}
}{a^2 H^2}\l( c_{e}^{2} \delta_{d}
 -\left(c_{e}^{2} -
c_{a}^{2}\right)\sigma_{d}\r) \label{Phi prime} \\
  \sigma^{\p}_{d} &=& \left( \l(\frac{\ddot{a}}{aH^2}\r) +3\left(c_{e}^{2} -
c_{a}^{2}\right)+ \l(3w
-1\r)\right)\frac{\sigma_{d}}{a}-\frac{3c_{e}^{2}\delta_{d}}{a}-\frac{3\Phi
(1+w)}{a}\label{sigma} \\
 \delta^{\p}_{d} &=&  \l(\frac{3\, \delta_{d}}{a}\r)\left(w - c_{e}^{2}\right) +
\frac{3\,\sigma_{d}}{a}\left(c_{e}^{2} -
c_{a}^{2}\right)+\frac{k^2\,\Phi}{3 a^3 H^2}
 + 3\left(1 + w\right)\Phi^{\p}.\label{delta d}
 \end{eqnarray}
 The quantities appearing in the above equations are as follows: i)  $\delta_{d}$ is the density contrast given by $\delta_{d} = \frac{\delta\rho_{d}}{\bar{\rho}_d}$, ii) $\sigma_d$ is a dimensionless variable  defined via equation  $\delta T^{0}_{~i}=\l(\rho_d/3H a^2\r)\sigma_{d,i}$, iii) $c_{a}^{2} \equiv \dot{\bar{p}}_{d}/\dot{\bar{\rho}}_{d}$ is the adiabatic sound speed of the dark energy. Here dot  and prime denotes the derivatives with respect to time and scale factor respectively.
Now solving the equations (\ref{Phi prime})-(\ref{delta d}) we obtain expressions for  $\Phi(a,\,k)$ and $\delta_{d}(a,\, k)$. Then, the density contrast $\delta_{m}$ describing the evolution of perturbations in the cold dark matter  can be evaluated from the
temporal component of the Linearized Einstein's equation and is given
by
\begin{eqnarray}\label{CDM pert}
 \delta_{m}(a,\, k)= -\l(\frac{2}{\Omega_{m}(a)}\r)\l(\Phi + a\, \Phi^{\p}+ \l(\frac{k^2 }{3H^{2}a^2}\r)\Phi\r)-\l(\frac{\Omega_{d}(a)}{\Omega_{m}(a)}\r)\,\delta_{d}(a,\, k),
\end{eqnarray}
~~~~~where $\Omega_{m}$ and $\Omega_{d}$ are the dimensionless
density parameter. It is evident from the equation (\ref{CDM pert}) that the matter power spectrum is dependent on the effective sound speed and equation of state. The corresponding expression for matter perturbation in the  $\Lambda$CDM will be independent of $c_e^2$ and $\delta_{d}$ as there are no perturbations to the cosmological constant. To see how the perturbations in dark energy effect the CDM power spectrum we will compare the growth of CDM perturbation in DE models with respect to $\Lambda$CDM
model.
\begin{figure}
\includegraphics[scale=0.350]{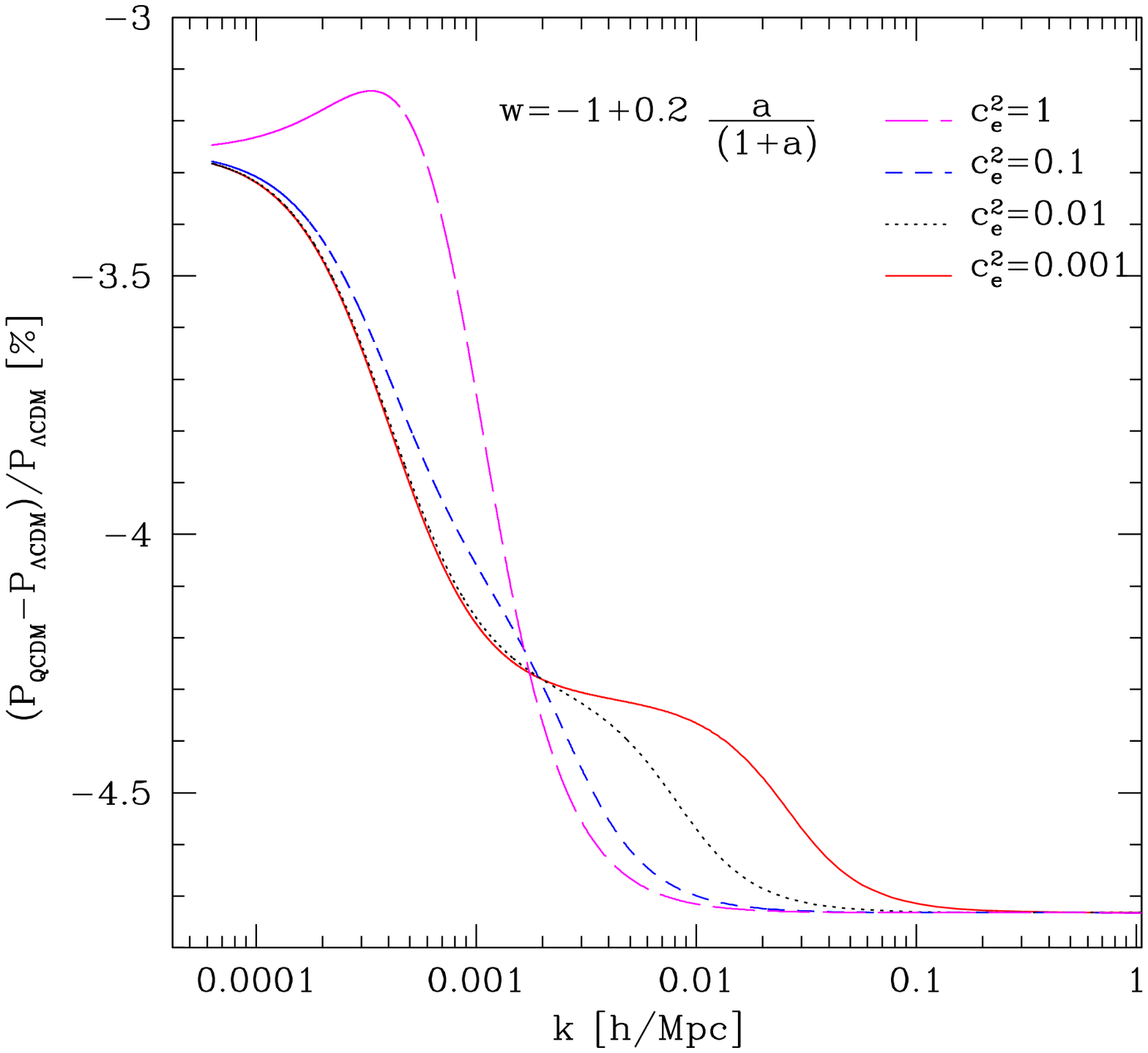}
\includegraphics[scale=0.35]{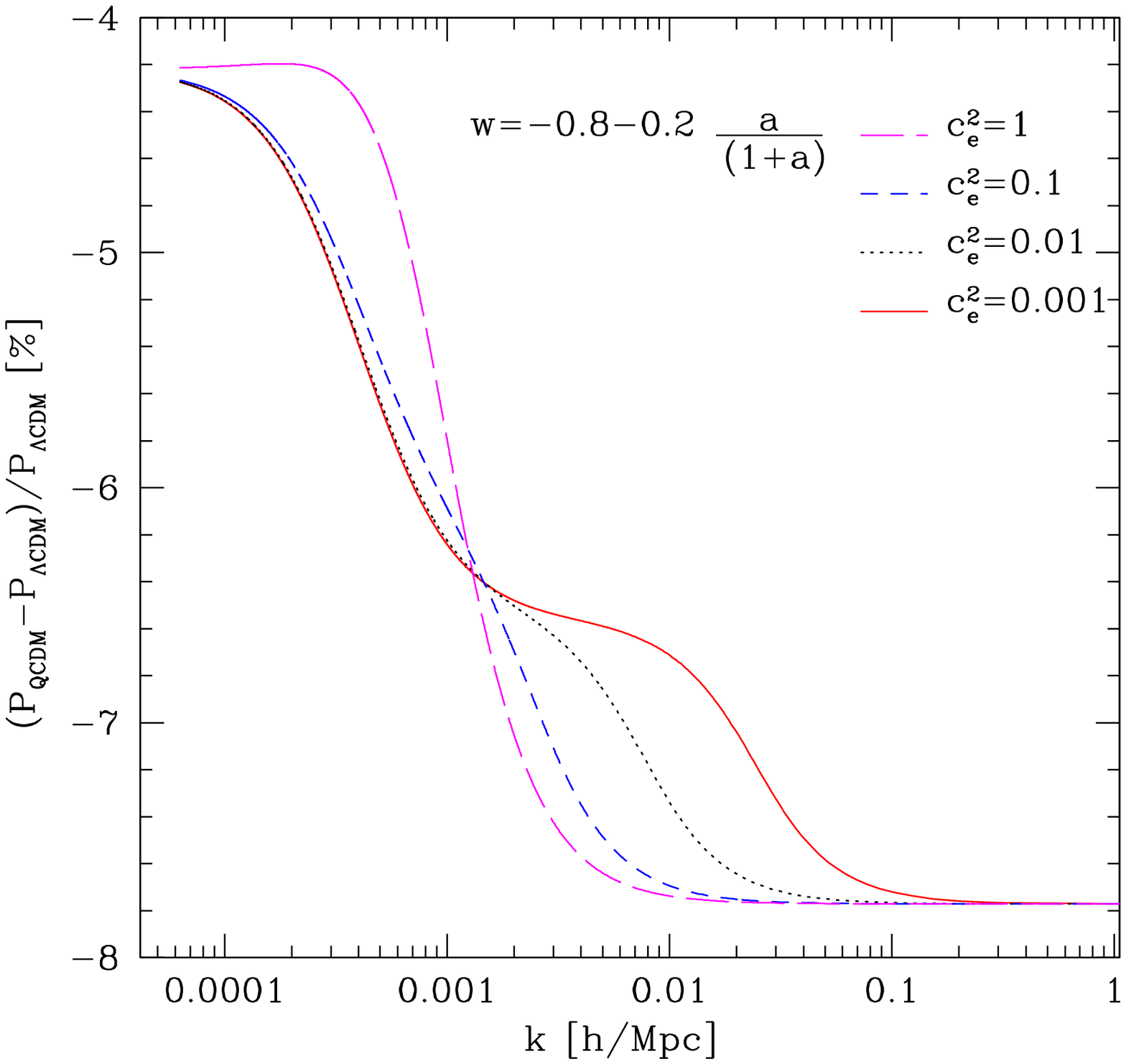}
\caption{Figure showing suppression in matter power spectrum
compared to $\Lambda$CDM model for the case when $w$ is evolving,
but $c_e^2$ is constant. In the left panel, $w$ is initially $-1$
and asymptotically approaches $-0.8$ whereas in the right panel
$w$ is initially $-0.8$ and asymptotically approaches $-1$. It can
also be verified that in both the panel $w$ at the present epoch
is $-0.9$.} \label{F2}
\end{figure}
\section*{Effect of sound speed and EOS parameter on CDM power spectrum}
We will now investigate, how varying the equation of state $w$ and sound speed $c_e^2$ of dark energy will influence CDM power spectrum. The most commonly studied scalar field dark energy models are with a constant value of $w$  and $c_e^2$. We consider
class of models in  which $c_e^2$ is constant and will  allow for
evolving EOS. The EOS $w$ can have a parametric form which depends
on the scale factor and can evolve from a initial value to
asymptotic value, spanning the entire evolution of the universe.
Then there are class of models where $c_e^2$ is epoch dependent and $w$ is
constant. Again we can use a similar parameterizations for $c_e^2$ which allows
for all possible values of $c_e^2$. Finally we can have the most
general case, where both $w$ and $c_e^2$ are epoch dependent (for
details of parameterization used see \cite{Rizwan}).
Next we quantitatively see how the sound speed $c_e^2$ and equation of state $w$ influences the behavior of power spectrum, by taking the difference in the growth of CDM perturbations in the scalar field models of dark energy with respect to the $\Lambda$CDM. The percentage difference can be found by numerically evaluating the following quantity
\begin{equation}\label{Delta}
\Delta \% \equiv \frac{P_{_{QCDM}}(k) - P_{_{\Lambda
CDM}}(k)}{P_{_{\Lambda CDM}}(k)}\times 100 \nn
\end{equation}
where $P_{_{QCDM}}(k)$ is the CDM power spectrum at the present
epoch in dark energy models whereas $P_{_{\Lambda CDM}}(k)$ is the
corresponding power spectrum in $\Lambda$CDM model.  Figure \ref{F1} shows
$\Delta$ for a constant value of $c_e^2$ and $w$, it can be seen that
at all length scales of perturbations percentage difference is negative. It can be inferred that the CDM power spectrum is suppressed in DE models in comparison to  $\Lambda$CDM model. Figure \ref{F2} shows the percentage difference in models where EOS is epoch dependent and $c_e^2$ is constant, again there is a suppression in CDM power spectrum at all length scales. Similarly, in the cases where  $c_e^2$ is evolving or
where both $c_e^2$ and $w$ are evolving with time, the CDM power spectrum is found to be suppressed compared to $\Lambda$CDM model \cite{Rizwan}.

There are certain key points to be noted from the Figures, the suppression depends on the value of $c_e^2$ and $w$. For instance a bigger value of equation of state, $w=-0.8$, gives more suppression than a smaller value, $w=-0.9$, implying the more EOS deviates from the value $\Lambda$CDM($w$ = -1) the suppression in CDM power spectrum increases. Also the effect of different values of $c_e^2$ (as seen in Fig \ref{F1} and \ref{F2}) is profound only at the intermediate length scales of $k$, whereas for scales much higher and smaller than the Hubble scale the value of $c_e^2$ is irrelevant. The reason why we see the effect of $c_e^2$ at intermediate scales is that the perturbation in dark energy grows only at scales greater than the sound
horizon (which for case of $c_e^2=0.001$ is $k = 0.07h/Mpc$, for other cases of $c_e^2$ considered here sound horizon will be greater than this value).
For scales much smaller than Hubble scale say  ($k >> 0.001h/Mpc$) the effect of $c_e^2$ is inappreciable, as the perturbation in
dark energy is negligibly smaller in comparison to that in dark matter. The dark energy can be approximated to be homogeneously distributed at these scales. On the other hand at scales much larger than the Hubble radius($k << 0.001h/Mpc$), it is the adiabatic perturbations of dark energy  which contribute to the CDM power spectrum, hence the effective speed of sound of dark energy does not influence
the CDM perturbations. Also, it turns out that the variation of an evolving $c_e^2$ does not have any considerable effect on cold dark matter power spectrum, whereas the  different evolution of $w$ will have much more significant influence \cite{Rizwan} . Hence exact information about the equation of state and speed of sound  will help in better understanding the nature of dark energy.

\end{document}